\documentclass{sigkddExp}
\usepackage{multirow}
\usepackage{float} 
\usepackage{url}

\begin{document}

\title{Stability and Controllability of Revenue Systems via the Bode Approach}

\numberofauthors{2}

\author{
\alignauthor Yichuan Niu\titlenote{Corresponding author email: yichuan.niu@utexas.edu}\\
\alignauthor Jianhui Chen\titlenote{The work was done while the author was with Snap Inc.}\\
}

\date{10 Feb. 2025}

\maketitle
\begin{abstract}
In online revenue systems, e.g. an advertising system, budget pacing plays a critical role in ensuring that the spend aligns with desired financial objectives. Pacing systems dynamically control the velocity of spending to balance auction intensity, traffic fluctuations, and other stochastic variables. Current industry practices rely heavily on trial-and-error approaches, often leading to inefficiencies and instability. This paper introduces a principled methodology rooted in Classical Control Theory to address these challenges. By modeling the pacing system as a linear time-invariant (LTI) proxy and leveraging compensator design techniques using Bode methodology, we derive a robust controller to minimize pacing errors and enhance stability. The proposed methodology is validated through simulation and tested by our in-house auction system, demonstrating superior performance in achieving precise budget allocation while maintaining resilience to traffic and auction dynamics. Our findings bridge the gap between traditional control theory and modern advertising systems in modeling, simulation, and validation, offering a scalable and systematic approach to budget pacing optimization.
\end{abstract}

\section{Introduction}
\vspace{4mm}

In most online revenue systems, monetary flows are controlled and optimized to achieve specific goals. For example, an advertising system's revenue flows from advertisers to opportunity suppliers (i.e., advertising platforms) are often dictated by the advertising products on advertiser's choices. Some advertisers prefer to distribute their budgets evenly throughout the day, while others may align spending with request traffic patterns. In this context, budget pacing control is essential to achieve their desired spending objectives.

\quad A pacing system creates a budget spending plan (referred to as the desired spend velocity) based on the advertising product being managed. This plan typically spans a single day or multiple days over the course of an advertising campaign. Equipped with one or more feedback loops, the pacing system monitors actual spending performance (the actual spend velocity) and actively adjusts it in near real-time to align with the spending plan. By minimizing the error between the desired and actual spend velocities, the system ensures targets are achieved \cite{pacing_2014, pacing_2015}.

\quad However, in such systems, designing budget pacing and monitoring actual spending performance can be challenging due to several factors \cite{niklas_long, niklas_short, niklas_2024}:

\vspace{-2mm}
\begin{itemize}
	\item Auction Dynamics: Fluctuations in ad retrieval, ranking, and competitive bidding strategies.
	\vspace{-2mm}
	\item Stochastic Traffic Behavior: Variability in user traffic and impression opportunities.
	\vspace{-2mm}
	\item Signal Noise: Discreteness in observed and desired spending signals.
	\vspace{-2mm}
	\item Environmental Variability: Changes in advertiser participants, auction intensity, budget constraints, and other time-dependent conditions.
\end{itemize}
\vspace{-2mm}

\quad In practice, pacing control is often developed by trial-and-error without a principled methodology for quantitatively addressing the stability, dynamics, and controllability of the system. Fortunately, while this presents a novel challenge for software engineers, many of these concerns can be addressed effectively using Classical Control Theory established for decades \cite{Ogata-control-textbook,Coughanower-control-textbook}.

\quad In this work, we apply Classical Control Theory to analyze and optimize online revenue systems, with a primary focus on advertising budget pacing. By modeling the pacing system as a linear time-invariant (LTI) proxy, we systematically derive compensators that improve stability and responsiveness while minimizing pacing errors. Using Bode analysis \cite{Franklin-control-textbook, control_textbook}, we design a robust control framework that ensures spending velocity remains within desired thresholds, even under dynamic auction conditions.
Our approach offers several key contributions: We formulate A control-theoretic framework for budget pacing, introducing mathematically grounded methods for stability and controllability analysis; we design and evaluate compensators, selecting optimal parameters meeting Bode stability criteria; The compensated system actively adapts to traffic fluctuations, bid competition, and auction intensity, ensuring consistent ad delivery and budget utilization; Our methodologies are validated through both simulated experiments and tested under real-world online auctions, demonstrating superior performance over the legacy step-based feedback control systems. 

\quad By bridging the gap between control theory and modern advertising platforms, this work introduces a scalable, systematic approach to budget pacing. Our findings highlight the advantages of compensator-based pacing over empirical tuning methods, providing a principled alternative for revenue optimization in dynamic, auction-driven environments.

\section{Related Work}
\vspace{4mm}

Budget pacing in online advertising is a fundamental problem that ensures advertisers maximize their return on investment while adhering to spending constraints. Approaches to budget pacing include heuristic-based methods, machine learning-driven optimization, and control theoretic frameworks. Below, we survey relevant literature across these key areas.

\quad Early budget pacing strategies relied on rule-based heuristics, such as uniform spending over time or proportional allocation based on past performance. For example, a proportional pacing algorithm is introduced in \cite{Feldman2006BudgetOI} that adjusts bid multipliers to control spending rates while maintaining competitiveness in ad auctions. Similar approaches have been widely used in industry due to their simplicity and interpretability \cite{Borgs2007DynamicsOB}. However, heuristic-based methods often lack robustness to dynamic market conditions.

\quad Recent research has leveraged machine learning for budget pacing by modeling demand and optimizing bid strategies. In \cite{bid-opt-Perlich}, Reinforcement learning is applied to dynamically adjust budgets in real-time bidding (RTB) environments. More recently, a deep reinforcement learning framework proposed in \cite{Jin2018RealTimeBW} learns optimal bidding strategies from historical auction data. These approaches improve efficiency but require extensive training data and may struggle with interpretability.

\quad While heuristic and machine learning-based methods have been widely adopted in industry, control-theoretic approaches provide a rigorous foundation for budget pacing. Niklas Karlsson has contributed significantly to the feedback control methodologies for online advertising from optimization point of view through years. Extensive research, simulation, and experiments have been conducted in his work \cite{niklas_long, niklas_short, niklas_2024, niklas_2022, niklas_2021, niklas_2022_simulation}. Similar study on RTB control can be found in Zhang's work \cite{zhang_2014, zhang_2016} as well as Yang's work \cite{yang_2019}.

\quad The existing literature lacks a connection between control theory and pacing performance in terms of stability and controllability via the Bode approach. Our work addresses this gap by leveraging classical control techniques for budget pacing, ensuring both stability and efficiency in online advertising environments.

\section{Real-time Revenue System}
\vspace{4mm}

To analyze and design a compensator for pacing, the ``plant" must first be modeled. Since the advertising revenue system exhibits time-varying, non-linear, stochastic, and discrete characteristics, we need to identify a linear time-invariant (LTI) proxy that is sufficiently simple to model yet capable of capturing the key dynamics.

\subsection{Bidding and Auction} 
\vspace{4mm}

\quad In an advertising system, the probability of conversion is often used to represent the likelihood of a desired event. In this work, we use the probability of an ad click as the conversion event. This probability ${p(click)}$ is provided in real-time by ad-ranking machine learning models based on the ad and the impression opportunity. Assuming ${p(click)}$ is perfectly calibrated, an ad will eventually receive a click when the cumulative summation of ${p(click)}$ across multiple impressions reaches $1.0$.

\quad To participate in an auction, advertisers must place a bid, ${max\_bid}$, representing the amount they are willing to pay for a click. For a single impression opportunity, each ad submits its final bid as follows \cite{nima_ads_opt, nima_auction_pacing}:

\begin{equation}
final\_ bid = max\_ bid~ \times p(click)
\end{equation}

\quad The ad with the highest bid is shown to the user, and the advertiser is charged either the final bid in a first-price auction or the runner-up's bid with second-price \cite{auction_textbook}. At this stage, the revenue system is not ``paced" because total ad spend is entirely dependent on auction traffic. In practice, however, advertisers operate within daily or lifetime budget constraints, ensuring that cumulative spending on all winning bids does not exceed these limits.

\subsection{Pacing}
\vspace{4mm}

To enforce such constraints, a pacing control variable is introduced into the final bid calculation:

\begin{equation}
paced\_ final\_ bid = \lambda \times ~final\_ bid
\end{equation}

where $\lambda \in (0,~1\rbrack$ acts as a bid modifier. This type of pacing is also known as discount pacing. The final auction and pricing process are then governed by the $\lambda$ stochastically.

\quad From bidding and auction to actual budget spending, the pacing control variable $\lambda$ influences the budget spent by rescaling bids, thereby increasing or decreasing the win rate in auctions. In most cases, $\lambda$ updates are scheduled at fixed intervals using a cron job.

\quad Under ideal conditions $\lambda$ remains constant as long as auction intensity and the total number of impression opportunities remain unchanged. However, in practice, $\lambda$ must adapt to environmental changes to ensure that the remaining budget is spent appropriately throughout the day or campaign lifetime, following the prescribed budget spending plan. The ultimate objective of this work is to design a suitable $\lambda$ controller, based on principles from Classical Control Theory in Section 5, to achieve precise and robust pacing.

\subsection{Spending Signal Sensing}
\vspace{4mm}

Spending signal sensing involves detecting the actual spending velocity for a specific pacing cohort while simultaneously performing denoising. Although it may appear unrelated to system performance, the sensing and smoothing algorithm plays a critical role in ensuring system responsiveness and stability because it interacts with the desired spending velocity in the feedback loop. The commonly used methods for signal sensing include (details in Appendix):

\vspace{-2mm}
\renewcommand{\labelitemi}{\hspace*{-1em}} 
\begin{itemize}
	\item \textbf{Exponential Smoother:} Exponentially smooth the signal by applying a weighted sum bewteen past values and current values.
	\vspace{-2mm}
	\item \textbf{Low-Pass Filter (LPF):} Canonical first-order or second-order filtering to reject high frequency noises.
\end{itemize}
\vspace{-2mm}

\quad Occasionally, moving average smoothing may be employed yet less favorable, since the moving window functions like a high-order integrator which often introduces significant lag into the control loop. In practice, we prefer a first-order low-pass filter, as it effectively removes most noise while maintaining sufficient responsiveness without a resonance frequency. Increasing the time constant $T_f$ provides stronger and smoother filtering yet introduces additional lag into the feedback loop.

\begin{equation}
LPF = \frac{1}{1 + sT_{f}},~~T_{f} = \frac{1}{2\pi f_{c}}~~
\end{equation}

where $s$ is the Laplace complex frequency variable which will be introduced in later sections. $f_c$ is the filter cutoff frequency which may be the inverse of minimum spending noise period in seconds. 

\quad More advanced yet uncommonly used filtering methods, such as the Kalman filter or G-H filter, require additional complexity of modeling the physical system and may also experience risks of diverging from the ground truth in some cases.

\subsection{The Plant}
\vspace{4mm}

To model the plant under control, i.e. the ad spending process, it is necessary to transform the discrete, non-linear, and time-varying processes of bidding, auction, and pricing into continuous linear time-invariant (LTI) components. This transformation can be done with following assumptions:

\textbf{1.} Auction occurs evenly and continuously producing a stream of ad spend.

\textbf{2.} The system operates in real-time from change of $\lambda$ to change of spend with negligible lag

\textbf{3.} At any given time mark, the relationship between $\lambda$ and budget spending velocity is monotonic and free of hysteresis.

\quad With these assumptions, the plant can be represented as a black box that provides a pointwise mapping from $\lambda$ to ad spending velocity. Assuming positive correlation between spending velocity and $\lambda$

\begin{equation}
sv = f\left(\Theta, \lambda \right)
\end{equation}

satisfy $sv \propto \lambda$, where $\Theta$ denotes other unknown factors relate to $sv$. At $\lambda = \lambda_n$, we have

\begin{equation}
sv(n) = \mathop{lim}\limits_{\Delta \lambda \rightarrow 0}\frac{f(\Theta ,\lambda_n) - f(\Theta ,\lambda_n + \Delta \lambda )}{\Delta \lambda } = W_n
\end{equation}

\quad This simplified modeling is only for controllability perspective, and our focus lies on the maximum $W_n$ and minimum $W_n$. If the controller can regulate the plant robustly within these bounds, the entire system will remain stable across its operational range. By analyzing historical auction data, the $W_n$ values corresponding to each $\lambda$-level can be determined. Typically, larger pacing cohorts (e.g., campaign level or ad account level) result in a smoother $\lambda$-to-spending velocity curve.

\quad While the discreteness of pricing and auctions can be ignored due to their high frequency, the discreteness introduced by pacing intervals cannot be neglected. There are two choices to model this interval: work out the whole design in sampling Z-domain or employ a signal holding mechanism. In this work, we opt for the later option by placing a Zero-Order Hold (ZOH) discretization at the input of the plant. The lag introduced by the ZOH mimics the pacing interval, adding a slight ``inertia" to an otherwise non-inertial plant. The system diagram is illustrated in Figure 1.

\begin{figure}[ht]
	\label{fig:system_diagram}
	\centering
	\includegraphics[scale=0.3]{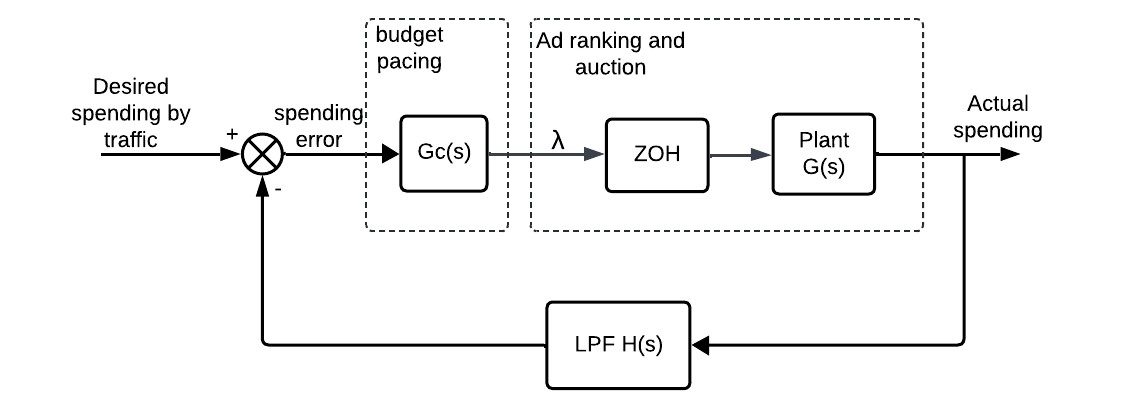}
	\caption{Feedback system diagram. $G_c(s)$ is compensator to be designed in later section, $G(s)$ is the modeled bidding and auction with $W_n$ gain, $H(s)$ is the LPF feedback path.}
\end{figure}

\section{Modeling Methodology}
\vspace{4mm}

\subsection{Laplace Transformation}
\vspace{4mm}

The Laplace transform is one of the most commonly used tools for analyzing and designing control systems. In the absence of a compensator, the open-loop transfer function of the ad ranking and auction system can be represented as the product of all components within the feedback loop.

\begin{equation}
ZOH \cdot G(s) \cdot H(s) = \frac{1 - e^{-sT_{ps}}}{s} \cdot W_n \cdot \frac{1}{1 + sT_f}
\end{equation}

\quad $T_{ps}$ represents the sampling rate of ZOH, which is defined by the pacing cycle. $T_f$ denotes the tunable decay parameter of the first-order LPF used to smooth the observed spending velocity. If the spend is measured at the impression level, $T_f$ may be smaller than $T_{ps}$. 

\quad To convert the system into a standard form for easier analysis, it is necessary to resolve the term $e^{- sT_{ps}}$. The Taylor expansion proves useful for this approximation, and we consider terms up to the 10th order.

\begin{equation}
e^{- sT_{ps}} = 1 - sT_{ps} + \frac{1}{2}\left( sT_{ps} \right)^{2}\cdots = \sum\limits_{n = 0}^{10}\frac{\left( - sT_{ps} \right)^{n}}{n!}
\end{equation}

\quad Note that choice of the approximation order affects the number of zeros in the system but does not influence the poles. The open-loop transfer function of the system is finally expressed as follows:

\begin{equation}
ZOH \cdot G(s) \cdot H(s) = \frac{W_{n}T_{ps}}{1 + sT_{f}}\sum\limits_{n = 2}^{10}\frac{\left( {- sT_{ps}} \right)^{n - 1}}{n!}
\end{equation}

\quad The plant has a single pole arising from the observed spending low-pass filter, which introduces a lag. Due to the high value of $W_n$,  typically resulting from the intensity of impression opportunities, the uncompensated plant becomes naturally unstable in response to the high-frequency (short-term) range due to the pole, meaning the ad spend will be diverging from the desired budget plan. Consequently, a compensator with at least one integrator is required to increase the phase margin and attenuate the gain contributed by $W_n$.

\subsection{Bode Method}
\vspace{4mm}

In classical control theory, several methods are used to evaluate system stability and dynamic performance, such as Nyquist stability criterion, root locus analysis, and Bode plot \cite{Ogata-control-textbook,Coughanower-control-textbook,Franklin-control-textbook, control_textbook}. In this work, we rely on the Bode stability criterion, specifically gain margin and phase margin, to quantitatively measure system stability. Additionally, the Bode method aids in compensator design from desired performance perspective, constructing compensators, and analyzing open-loop and closed-loop stability and dynamics, offering more intuitive visualization compared to other methods.

\vspace{-2mm}
\renewcommand{\labelitemi}{\hspace*{-1em}} 
\begin{itemize}
	\item \textbf{Gain margin (GM):} The gain margin measures how much the system gain can be increased before it becomes unstable. A larger gain margin indicates a more robust system capable of tolerating variations in gain. In our context, ``robust" refers to the system's resilience to fluctuations in request traffic and auction intensity.
	\vspace{-2mm}
	\item \textbf{Phase margin (PM):} The phase margin represents the additional input-to-output lag that can be introduced into the ad system before instability occurs. A larger phase margin indicates a more stable system with reduced oscillatory behavior. Understanding phase margin is critical for assessing how spending noise affects stability.
\end{itemize}
\vspace{-2mm}

\quad By analyzing the Bode plot of the open-loop plant in Figure 2, we observe that both the phase margin and gain margin are negative, indicating instability in the absence of a compensator. The high gain further reflects an extremely sensitive $\lambda$-to-spending response, emphasizing the need for stabilization.

\begin{figure}[ht]
	\label{fig:plant_bode}
	\centering
	\includegraphics[scale=0.45]{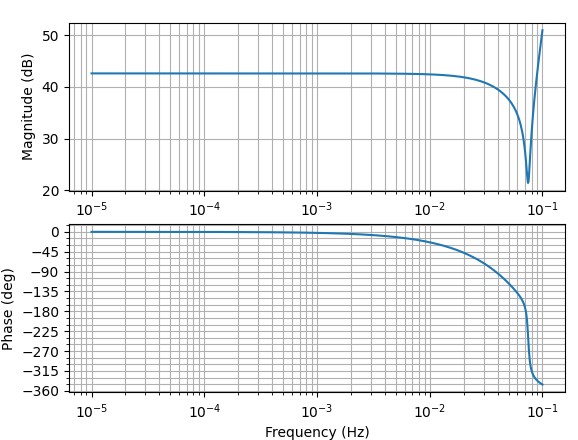}
	\caption{Bode plot of the open-loop transfer function. $W_n = 13.52$, $T_{ps} = 10$, $T_f = 2 / 2\pi$, aliasing effects for $f > 5 \times 10^{- 2}$ Hz.}
\end{figure}

\quad The system gain dropped drastically above $5 \times 10^{- 2}$ Hz. This behavior is consistent with the Nyquist–Shannon sampling theorem, which states that for a sampling period of $T_{ps}=10$ seconds, the maximum frequency for lossless information representation must be below $5 \times 10^{- 2}$ Hz. Any gain increase observed beyond $1 \times 10^{- 1}$ Hz is unreliable due to aliasing effects, which shall be ignored.

\section{Compensator Design}
\vspace{4mm}

With the modeled plant, the compensator $G_{c}(s)$ will be designed in the Laplace domain using Bode analysis. The design philosophy is twofold: The compensated system must remain stable when $W_n$ is high during intense auction states. Additionally, the system must exhibit sufficient dynamics when $W_n$ is low; otherwise, it will respond too slowly to mild auction states.

\quad There are two main schools of compensator design methodologies:

\begin{itemize}
	\item \textbf{School of Practitioner:} This approach starts with a predefined compensator, such as a Propotional controller, based on the plant's characteristics. The compensated open-loop transfer function is derived, and the compensator parameters are tuned to achieve stability and adequate dynamic performance.
	\vspace{-2mm}
	\item \textbf{School of Theorist:} This approach begins with a desired steady-state and frequency response, as visualized on the Bode plot. A parametrized compensator transfer function is then numerically composed by placing zeros and poles to fit the desired Bode response. Finally, the compensator transfer function is converted from the Laplace domain to the Z-domain, and subsequently into a time-domain recurrence relation for programming implementation.
\end{itemize}

Based on the Bode plot, the desired compensator for our plant must exhibit the following characteristics:

\quad \textbf{1.} Quench amplification: The high gain introduced by auction intensity ($W_n$) creates excessive sensitivity that must be mitigated to ensure stability.

\quad \textbf{2.} Increase phase margin:At the 0 dB crossover frequency, the phase margin (relative to $180^{\circ}$ ) must be increased from negative to positive to reduce overshoot and oscillation caused by noises.

\subsection{Practitioner Methodology: PID Compensator}
\vspace{4mm}

With the practitioner approach, a compensator is preselected, and its parameters are adjusted to align with the plant's stability and dynamic performance requirements. Among compensators, the Proportional–Integral–Derivative (PID) controller is arguably the most widely used due to its simplicity and effectiveness \cite{control_textbook}.

\begin{equation}
G_c(s) = K_p + \frac{K_i}{s} + K_ds
\end{equation}

\quad The plant exhibits low inertia and high gain, making the differential term $K_{d}s$ potentially hazardous, as it would further amplify the gain at high frequencies. Staying with P, and I terms, the best $K_p$ and $K_i$ can be determined through a grid search, balancing GM and PM considerations. Figure 3 shows the PI compensator in the Bode plot. The characteristics of lowing gain and increasing phase angle at high frequency is desirable for our plant.

\begin{figure}[ht]
	\label{fig:pi_compensator_bode}
	\centering
	\includegraphics[scale=0.38]{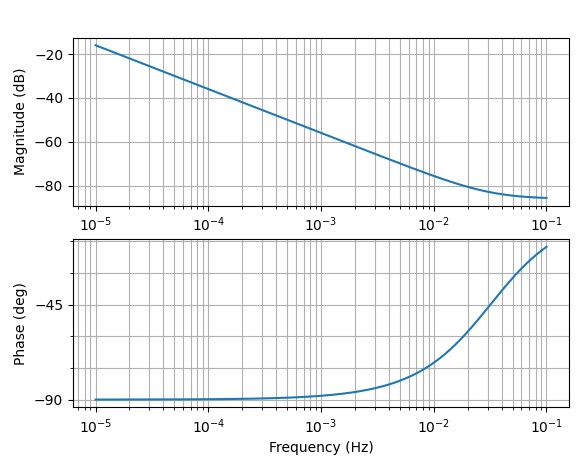}
	\caption{PI compensator Bode plot. $K_p = 5 \times 10^{-5}$, $K_i = 1 \times 10^{-5}$.}
\end{figure}

\subsection{Theorist Methodology: Zero-Pole Compensator}
\vspace{4mm}

A more mathematical approach to design a compensator is to define a generic compensator with zeros and poles. The formulation of the generic compensator is as follows \cite{control_textbook}:

\begin{equation}
G_{c}(s) = K_{c}\frac{\left( {\frac{s}{z_{1}} + 1} \right)\left( {\frac{s}{z_{2}} + 1} \right)~\cdots}{\left( {\frac{s}{p_{1}} + 1} \right)\left( {\frac{s}{p_{2}} + 1} \right)~\cdots}
\end{equation}

where $K_c$ is the compensator global gain, bringing $+ 20log_{10}K_{c}$ dB lift with no change in phase angle in the Bode plot. $z_1$, $z_2$, $\cdots$ are the locations of zeros and $p_1$, $p_2$, $\cdots$ are the locations of poles. Each zero brings a $+20$ dB/decade slope starting at $f = z_i / 2\pi$ Hz with a $90^{\circ}$ leading phase at $[z_i - decade, z_i + decade]$, whereas each pole decays gain by a $-20$ dB/decade slope starting at $f = p_i / 2\pi$ Hz with a $90^{\circ}$ lagging phase at $[p_i - decade, p_i + decade]$ on the Bode plot.

\begin{figure}[ht]
	\label{fig:zp_compensator_bode}
	\centering
	\includegraphics[scale=0.39]{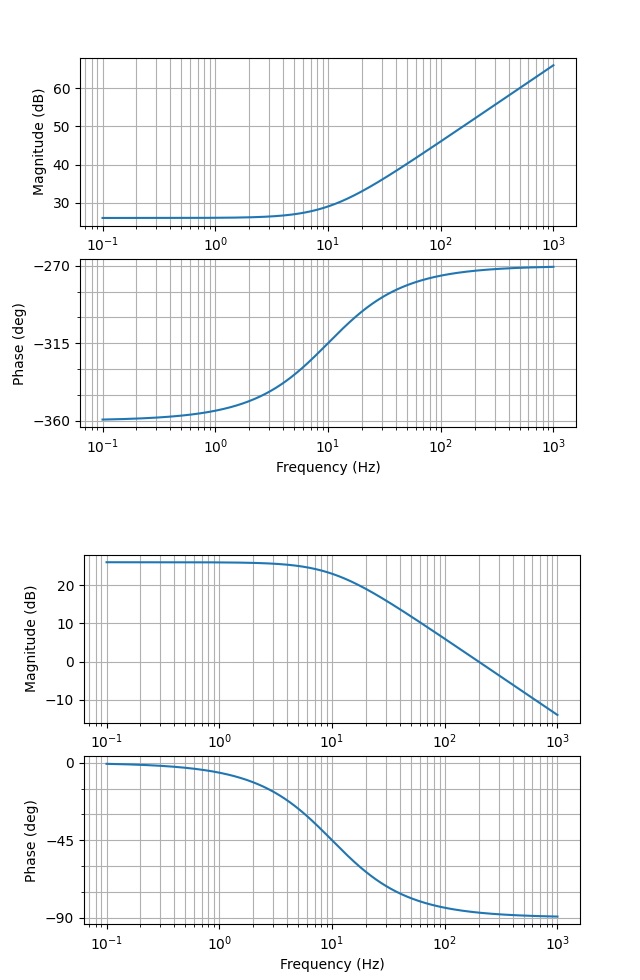}
	\caption{Upper: Zero with $K_c = 20, z_1 = 10 \times 2\pi $. Lower: Pole with $K_c = 20, p_1 = 10 \times 2\pi$}
\end{figure}

\quad Therefore, the compensator can be crafted using the building components $K_c$, $z_i$, and $p_i$ to shape the final system transfer function and achieve the desired Bode curve.

\quad Note that once the generic compensator has been parameterized, there is still no straightforward method for implementing it in computer programming. To overcome this, the compensator transfer function must be converted from the continuous Laplace domain to the discrete Z-domain using the Tustin (Bilinear) Approximation. This involves substituting the frequency variable s with:

\begin{equation}
s = \frac{2}{T_z}\frac{1 - Z^{- 1}}{1 + Z^{- 1}}
\end{equation}

where $T_z$ is the sampling period that could be the same as the plant $T_{ps}$. By replacing $G_{c}(s)$ with the discrete output-to-input ratio at step $k$, $y[k] / u[k]$, we may further convert Z-domain representation into a recurrence relation of the following form:

\begin{equation}
\begin{array}{l}
{y\lbrack k\rbrack = a_{1}y\left\lbrack {k - 1} \right\rbrack + a_{2}y\left\lbrack {k - 2} \right\rbrack + \cdots + b_{0}u\lbrack k\rbrack + b_{0}u\left\lbrack {k - 1} \right\rbrack + \cdots} \\
{= \sum\limits_{i = 1}^{m}a_{i}y\left\lbrack {k - i} \right\rbrack + \sum\limits_{i = 0}^{n}b_{i}u\left\lbrack {k - i} \right\rbrack}
\end{array}~~~
\end{equation}

\quad Programming this recurrence relation, with memorization of previous inputs and outputs, can ultimately implement the desired compensator in a digital system.

\section{Experiments}
\vspace{4mm}

\subsection{Evaluation Metrics}
\vspace{4mm}

\begin{table*}[ht]
	\centering
	\caption{Compensator Parameters vs GM, PM, cutoff gain (COG), and closed-loop bandwidth (CL-BW).}
	\begin{tabular}{cccccccccc}
		\cline{3-10}
		& \multicolumn{1}{c|}{}           & \multicolumn{4}{c|}{Max $W_n$}                                                            & \multicolumn{4}{c|}{Min $W_n$}                                                          \\ \hline
		\multicolumn{1}{|c}{$K_p$}    & \multicolumn{1}{c|}{$K_i$}         & GM (dB)              & PM ($^{\circ}$)   & COG (dB)         & \multicolumn{1}{c|}{CL-BW (Hz)}          & GM (dB)              & PM ($^{\circ}$)          & COG (dB)         & \multicolumn{1}{c|}{CL-BW (Hz)} \\ \hline
		\multicolumn{1}{|c}{5e-2}  & \multicolumn{1}{c|}{5e-3}       & -2.67                & -13.05                  & -1.445e-05               & \multicolumn{1}{c|}{7.51e-02}                & 19.69                & 98.25                    & -3.175e-4                & \multicolumn{1}{c|}{1.75e-02}         \\ \hline
		\multicolumn{1}{|c}{5e-3}  & \multicolumn{1}{c|}{5e-3}       & -3.83                & -18.58                  & 2.506e-05                & \multicolumn{1}{c|}{6.70e-02}                & 18.54                & 73.72                    & -5.141e-6                & \multicolumn{1}{c|}{2.50e-02}         \\ \hline
		\multicolumn{1}{|c}{5e-4}  & \multicolumn{1}{c|}{5e-3}       & -5.91                & -33.28                  & 2.901e-05                & \multicolumn{1}{c|}{6.47e-02}                & 16.45                & 71.12                    & -2.610e-5                & \multicolumn{1}{c|}{2.64e-02}         \\ \hline
		\multicolumn{1}{|c}{5e-2}  & \multicolumn{1}{c|}{5e-4}       & -0.1                 & -0.55                   & -4.404e-3                & \multicolumn{1}{c|}{7.57e-02}                & 22.26                & 118.72                   & -5.222e-02               & \multicolumn{1}{c|}{9.94e-04}        \\ \hline
		\multicolumn{1}{|c}{5e-3}  & \multicolumn{1}{c|}{5e-4}       & 17.33                & 98.29                   & -4.705e-4                & \multicolumn{1}{c|}{1.12e-02}                & 39.69                & 91.0                     & -2.142e-02               & \multicolumn{1}{c|}{1.33e-03}        \\ \hline
		\multicolumn{1}{|c}{5e-4}  & \multicolumn{1}{c|}{5e-4}       & 16.17                & 68.82                   & -7.54e-05                & \multicolumn{1}{c|}{1.79e-02}                & 38.54                & 88.35                    & -1.831e-02               & \multicolumn{1}{c|}{1.43e-03}        \\ \hline
		\multicolumn{1}{|c}{5e-2}  & \multicolumn{1}{c|}{5e-5}       & 0.11                 & 0.61                    & \textbf{-3.359e-01}      & \multicolumn{1}{c|}{7.59e-02}                & 22.47                & \textbf{120.77}          & \textbf{-2.878}          & \multicolumn{1}{c|}{9.68e-05}        \\ \hline
		\multicolumn{1}{|c}{5e-3}  & \multicolumn{1}{c|}{5e-5}       & 19.9                 & \textbf{129.05}         & -7.261e-02               & \multicolumn{1}{c|}{7.99e-04}                & 42.26                & 92.76                    & -1.876                   & \multicolumn{1}{c|}{1.27e-04}        \\ \hline
		\multicolumn{1}{|c}{5e-4}  & \multicolumn{1}{c|}{5e-5}       & 37.33       & 91.32                   & -3.383e-02               & \multicolumn{1}{c|}{1.05e-03}                & 59.69       & 90.1                     & -1.673                   & \multicolumn{1}{c|}{1.35e-04}        \\ \hline
		\multicolumn{1}{l}{}       & \multicolumn{1}{l}{}            & \multicolumn{1}{l}{} & \multicolumn{1}{l}{}    & \multicolumn{1}{l}{}     & \multicolumn{1}{l}{}                         & \multicolumn{1}{l}{} & \multicolumn{1}{l}{}     & \multicolumn{1}{l}{}     & \multicolumn{1}{l}{}             \\ \hline
		\multicolumn{1}{|c}{Zeros} & \multicolumn{1}{c|}{Poles}      & GM (dB)              & PM ($^{\circ}$)         & COG (dB)         & \multicolumn{1}{c|}{CL-BW (Hz)}          & GM (dB)              & PM ($^{\circ}$)          & COG (dB)         & \multicolumn{1}{c|}{CL-BW (Hz)} \\ \hline
		\multicolumn{1}{|c}{1e-1}  & \multicolumn{1}{c|}{1e-4, 1e-3} & \textbf{60.52}                & 17.77                   & 1.456e-01                & \multicolumn{1}{c|}{8.98e-04}                & \textbf{82.89}                & 27.81                    & 5.541e-01                & \multicolumn{1}{c|}{2.92e-04}         \\ \hline
		\multicolumn{1}{|c}{1e-1}  & \multicolumn{1}{c|}{1e-4}       & 31.31                & 93.03                   & -7.273e-02               & \multicolumn{1}{c|}{2.09e-03}                & 53.67                & 95.77                    & -9.292e-01               & \multicolumn{1}{c|}{2.86e-04}        \\ \hline
		\multicolumn{1}{|c}{1e-1}  & \multicolumn{1}{c|}{1e-3}       & 11.34                & 63.06                   & -6.416e-02               & \multicolumn{1}{c|}{6.63e-02}                & 33.7                 & 97.54                    & -4.987e-01               & \multicolumn{1}{c|}{2.76e-03}        \\ \hline
	\end{tabular}
\end{table*}

\begin{figure*}[!ht]
	\label{fig:ki_kp_gm_pm}
	\centering
	\includegraphics[scale=0.45]{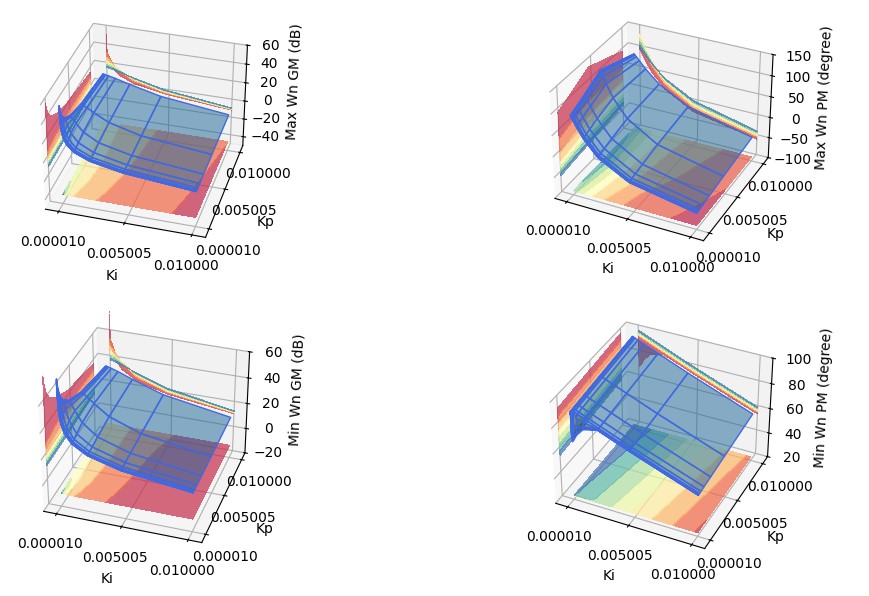}
	\caption{$K_p$ and $K_i$ vs GM and PM regarding maximum $W_n$ and minimum $W_n$}. Positive GM and PM at maximum $W_n$ with largest PM at minimum $W_n$ is preferred.
\end{figure*}

To evaluate the performance of the proposed compensator, we use the stability and dynamic criteria outlined in Section 4.2. Additionally, we adopted the pacing error metrics from \cite{ebay_pacing} measuring the normalized system pacing error through simulation. The pacing error is defined as:

\begin{equation}
PE = \frac{1}{N}\sum_{i = 1}^N\left(\frac{1}{M}\sum_{j = 1}^M\frac{\left|dS_{ij} - aS_{ij}\right|}{dS_{ij}}\right)
\end{equation}

where $N$ is the number of pacing cohorts, $M$ is the number of measurements, $dS_{ij}$ is the desired spending velocity for cohort $i$ at pacing period $j$ and $aS_{ij}$ is observed actual spending velocity. $PE$ represents the total relative execution error from the desired spending velocity. 

\quad From a business perspective, we can further weight $PE$ by the cohort spend as follows \cite{ebay_pacing}:

\begin{equation}
\begin{matrix}
{SWPE = \frac{1}{N}\sum\limits_{i = 1}^{N}W_{i} \cdot \left( {\frac{1}{M}\sum\limits_{j = 1}^{M}\frac{\left| {dS_{ij} - aS_{ij}} \right|}{dS_{ij}}} \right)} \\
{subject~to~\sum\limits_{i = 1}^{N}W_{i} = 1~~~~}
\end{matrix}
\end{equation}

where $W_i$ represents the cohort's daily spend weight.

\subsection{Simulation of the Compensated Plant}
\vspace{4mm}

The compensator and ad system, initially represented in the Laplace domain, are converted and simulated in the time domain (top graph of Figure 7). Key simulation parameters (Ad Set 1 in Appendix) are derived from our in-house data:

\begin{itemize}
	\item $max(W_n) = 13.52$ \$/($\lambda \cdot$ minute)
	\vspace{-2.5mm}
	\item $min(W_n) = 1.707$ \$/($\lambda \cdot$ minute)
	\vspace{-2.5mm}
	\item $T_{ps} = 10$ second
	\vspace{-2.5mm}
	\item $T_{as} = 0.87$ second
	\vspace{-2.5mm}
	\item $T_{f} = 10/2\pi$ second
	\vspace{-2.5mm}
	\item Daily budget: \$ $387.5$
	\vspace{-2.5mm}
	\item Traffic curve pattern: Normalized ads request curve per ad set
	\vspace{-2.5mm}
	\item Desired spending velocity: $\frac{remaining\_ budget \times current\_ traffic}{remaining\_ traffic}$
\end{itemize}

\quad To be more realistic, random noise is drawn from a Gaussian distribution with a mean equal to $G(s)$ output and a variance equal to $5$\% of the output value, which is then added to the actual spending velocity. The maximum $W_n$ represents the lower bound for stability, while the largest PM at the minimum $W_n$ provides the best dynamics. An ideal compensator should exhibit a positive GM and PM for maximum $W_n$, as well as the largest PM for the minimum $W_n$. 

\quad \textbf{PI compensator:} In the compensated pacing system, the open-loop transfer function is given by $G_{c}(s)\cdot ZOH\cdot G(s)\cdot H(s)$.  Since P control alone is insufficient to stabilize the system due to missing high-frequency noise rejection, and D control is undesirable as it amplifies feedback noises and thereby reducing the signal-to-noise (S/N) ratio. In the Practitioner method, we select a PI compensator and perform a grid search for parameters that satisfy the Bode stability criteria. To prevent integrator windup, the integrator is limited to the range $[0, 0.5]$ and is only activated when the lumped output is within the range $(0, 1)$. Nine combinations of $K_p$ and $K_i$ with their stability performance can be found in Table 1. We can see that a stronger integrator (larger $K_i$) must match with a higher gain (larger $K_p$) to maintain stability. However, $K_p$ will find its limitation to stabilize the system once $K_i$ goes beyond a certain point. Their corresponding plots are shown in Figure 5. 

\quad \textbf{Zero-pole compensator:} As a demonstration, three zero-pole compensators are designed following the methodologies outlined in Section 5.2 with $K_c = 1.0$ In these compensators, zeros capture the input error history, while poles capture the output ($\lambda$) history. Therefore, adding more zeros improves the dynamics with respect to the spending error, whereas poles help maintain steady pacing and slow down the response. We can artificially create zeros to cancel the poles from the ad system modeling and improve dynamics. Also found in Table 1, there are three examples of zero-pole compensators. Poles trade PM for GM while zeros behaves vice versa.

\begin{figure}[H]
	\label{fig:input_fft}
	\centering
	\includegraphics[scale=0.4]{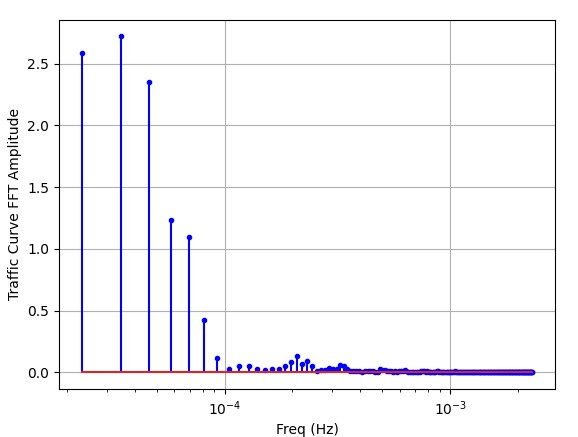}
	\caption{Traffic curve FFT decomposition. Since the system is supposed to track the traffic curve as desried spending, the maximum frequency of the FFT spectrum shall be below the system closed-loop bandwidth.}
\end{figure}

\quad Additionally, we measure the traffic curve spectrum using the Fast Fourier Transform (FFT) in Figure 6. Since pacing functions as a tracking system, we expect the closed-loop system to have a negative gain for frequencies above the highest traffic spectrum at $9.3 \times 10^{-5}$ Hz, indicating resilience to noise. For the compensators tested Table 1, the lowest closed-loop bandwidth is $9.68 \times 10^{-5}$ Hz, which is sufficient to adapt the traffic fluctuation pattern.

\subsection{Real-World Testing}
\vspace{4mm}

\quad To evaluate the validity of the designed compensator, we tested seven top-spending ad sets (listed in the Appendix) from our Click ad product. These ad sets were selected from the 1/20 budget split group and tested for one day on January 12, 2025 (PT). Due to the short testing period, we preloaded the integrator by an initial $\lambda$ from the last value prior to the test to ensure the system reached equilibrium faster, mimicking continuous operation. Each ad set was paced according to its targeted daily traffic curve by its own independent $\lambda$ compensator. The pricing employed a second-price auction model following the legacy business logic.

\quad In Figure 7, we compare time-domain simulated spending curve and real-world auction spending curve from our methodologies against the legacy pacing system, which employs a step-based control mechanism. The simulated pacing closely aligns with real-world auction outcomes, supporting the effectiveness of the control system modeling. Our method demonstrates a more stable spending velocity, with reduced overshoot and undershoot compared to the legacy system. PE and SWPE comparisons are both improved dramatically in Table 2. Figure 8 shows the control variables and signals of Ad Set 1 during the 1-day auction. 

\begin{table}[!h]
	\centering
	\caption{PE and SWPE comparison to legacy pacing.}
	\begin{tabular}{ccc}
		\multicolumn{1}{l}{} & PE     & SWPE    \\ \hline \hline
		Ours                 & 0.1650 & 0.01741 \\ \hline
		Legacy Pacing   & 0.3731 & 0.03109 \\ \hline \hline
	\end{tabular}
\end{table}

\begin{figure}[ht]
	\label{fig:sim_auction_legacy_comparison}
	\centering
	\includegraphics[scale=0.38]{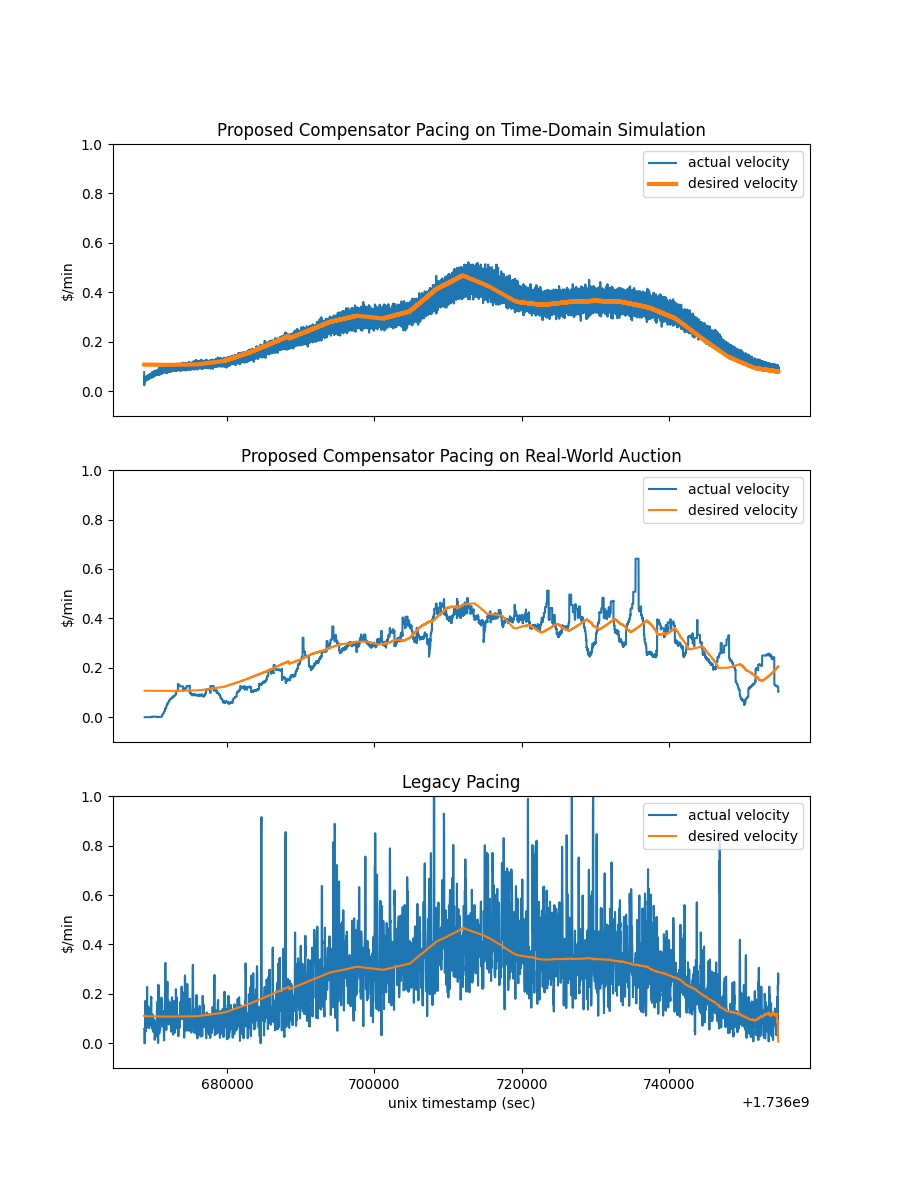}
	\caption{spending curve comparison of Ad set 1. Upper: Time-domain simulated system; Middle: Real-world auction; Lower: Legacy pacing.}
\end{figure}

\begin{figure}[ht]
	\label{fig:auction_signals}
	\centering
	\includegraphics[scale=0.33]{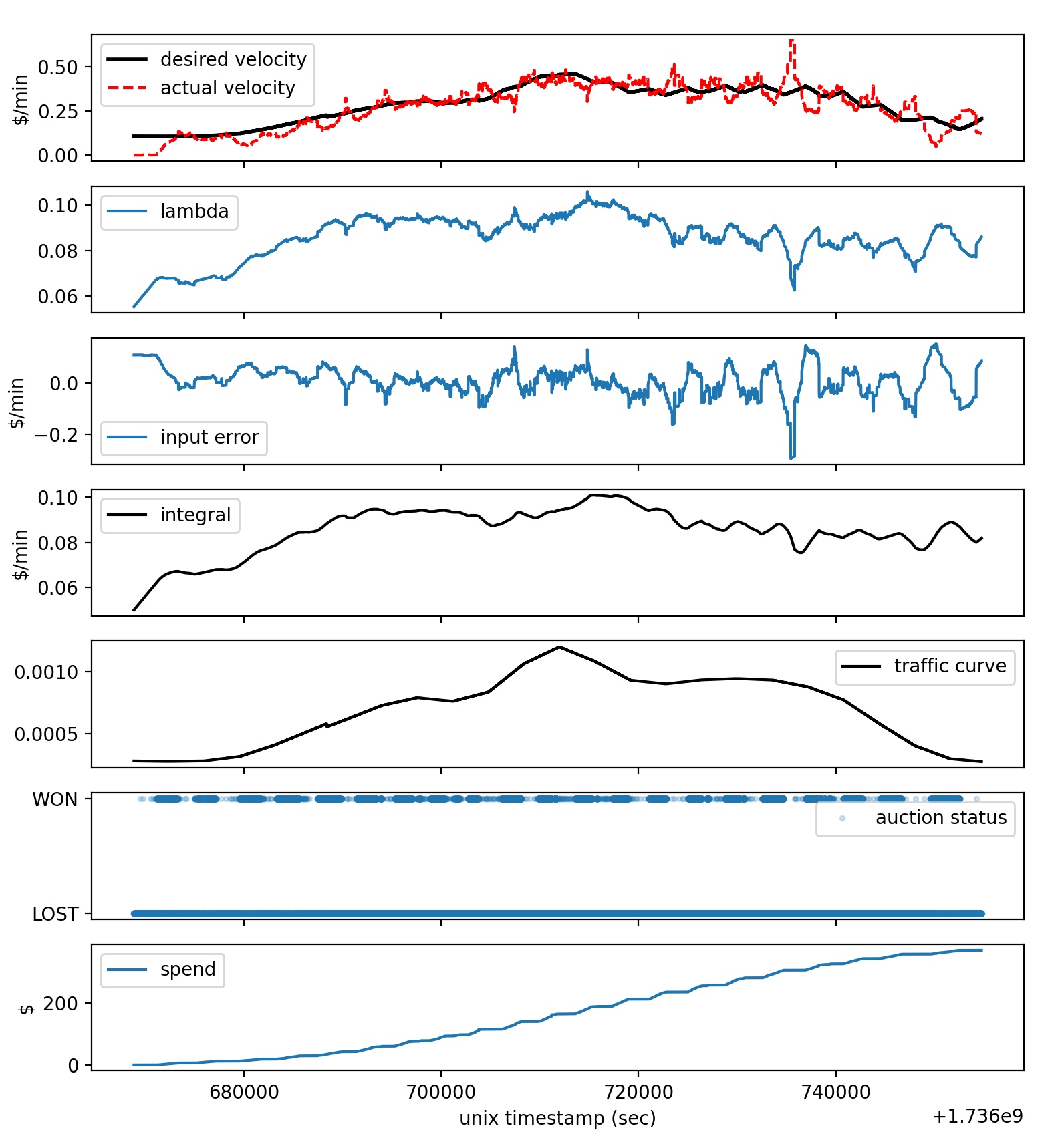}
	\caption{More signals of Ad set 1 from real-world auction.}
\end{figure}

\section{Conclusion and Future Work}
\vspace{4mm}

In this work, we demonstrate system modeling and compensator design for budget pacing in a revenue system by treating the ad ranking, bidding, and auction system as a LTI system. Simulations and real-world auctions show significant improvements compared to our legacy system. To enhance compensator flexibility, our future work aims to pursue a time-varying compensator design in future iterations. The compensator may utilize multiple sets of parameters. Upon detecting a change, the compensator may switch between presets for improved adaptability. Additionally, we observe that $\lambda$ is not always positively correlated with spending velocity due to fluctuations in impression opportunities and auction intensity. This issue could be addressed by modeling an additional feedforward path as a correction factor.

\bibliographystyle{unsrt}
\bibliography{kdd_refs}  

\balancecolumns

\appendix

\section{Low Pass Filter for Observing Actual Spending Velocity}
\vspace{4mm}

\subsection{Canonical 1st Order LPF}
\vspace{4mm}
In Section 3.3, we introduced a first-order LPF to reduce noise in measuring the actual spending velocity. Since the filter is defined in the Laplace domain, we now derive its time-domain implementation. The LPF transfer function is given by:

\begin{equation}
H(s) = \frac{1}{1 + T_{f}s}
\end{equation}

\quad It can be converted into the Z-domain by applying Tustin’s transformation, as discussed in Section 5.2. The resulting LPF in the Z-domain is given by:

\begin{equation}
\begin{array}{r}
{H(z) = \frac{1}{1 + T_{f}\frac{2}{T}\frac{1 - z^{- 1}}{1 + z^{- 1}}}} \\
{= \frac{\frac{T}{T + 2T_{f}}\left( 1 + z^{- 1} \right)}{1 + \frac{T - 2T_{f}}{T + 2T_{f}}z^{- 1}}}
\end{array}
\end{equation}

where $T$ represents the sampling period of the LPF. In an ad auction system, it corresponds to the auction-winning and pricing interval of the paced cohort. Since $T$ is inherently variable, we approximate it by quantizing the auction interval to its median and imputing missing values using linear interpolation. For mathematical convenience, we define $a$ and $b$ as follows:

\begin{equation}
\begin{matrix}
{a = \frac{T - 2T_{f}}{T + 2T_{f}},} & {b = \frac{T}{T + 2T_{f}}}
\end{matrix}
\end{equation}

\quad The output-to-input relationship is therefore as follows:

\begin{equation}
\frac{Y(z)}{U(z)} = \frac{b\left( 1 + z^{- 1} \right)}{1 + az^{- 1}}
\end{equation}

\quad Expand the equation:

\begin{equation}
Y(z) + aY(z)z^{- 1} = bU(z) + bU(z)z^{- 1}
\end{equation}

\quad Converting the Z-domain representation into a time-domain recurrence relation, the final first-order LPF difference equation for implementation is given by:

\begin{equation}
y\lbrack k\rbrack = bu\lbrack k\rbrack + bu\lbrack k - 1\rbrack - ay\lbrack k - 1\rbrack
\end{equation}

\subsection{Exponential Smoother}
\vspace{4mm}
In practice, we’ve found that engineers rarely use the canonical first-order LPF, as its time-domain formulation is not intuitive. Instead, software engineers commonly prefer the exponential decay method for smoothing, as shown below:

\begin{equation}
y\lbrack k\rbrack = \beta u\lbrack k\rbrack + (1 - \beta)y\lbrack k - 1\rbrack
\end{equation}

where $\beta$ is a weighting parameter usually in range $\beta \in (0,1)$. To address this use case, we derive the transfer function in the Laplace domain, enabling it to replace the canonical LPF for system stability and performance analysis. We then convert the corresponding time-domain recurrence relation into the Z-domain and express it as a transfer function.

\begin{equation}
H(z) = \frac{Y(z)}{U(z)} = \frac{\beta}{1 - (1 - \beta)z^{- 1}}
\end{equation}

\quad Use inverse Tustin's transformation $z = (1 + 0.5sT)~/~(1 - 0.5sT)$, where $T$ represents the same ad auction winning interval. Ultimately, the transfer function of the exponential smoother in the Laplace domain is given by:

\begin{equation}
H(s) = \frac{\beta(1 + 0.5sT)}{(1 - 0.5\beta)sT + \beta}
\end{equation}

%
%
%
%
%

\section{Tested Ad Sets Info}
\vspace{4mm}

The ad sets under test in Section 6.3 are configured as follows. They are daily budget ad sets from a $1/20$ budget split group. Each ad set owns a indenpendent pacing $\lambda$. Their initial $\lambda$ preloaded for the integrator is the last value before the test.

\begin{table}[ht]
	\begin{tabular}{rcc}
		& Daily Budget (\$) & Initial $\lambda$ \\ \hline \hline
		Ad Set 1 & 387.5             & 0.05      \\ \hline
		Ad Set 2 & 250               & 0.2       \\ \hline
		Ad Set 3 & 800               & 0.015     \\ \hline
		Ad Set 4 & 500               & 0.02      \\ \hline
		Ad Set 5 & 111               & 0.07      \\ \hline
		Ad Set 6 & 275               & 0.017     \\ \hline
		Ad Set 7 & 248               & 0.5       \\ \hline \hline
	\end{tabular}
\end{table}

\end{document}